%% file: OT_longit_PRD.tex
\documentclass[%
 reprint,
 superscriptaddress,
showpacs,
 amsmath,amssymb,
 aps,
 prd,
]{revtex4-1}

\usepackage{graphicx}% Include figure files
\usepackage{dcolumn}% Align table columns on decimal point
\usepackage{bm}% bold math
\usepackage{color}
\usepackage{dsfont}
\usepackage{acronym}
\acrodef{eom}[EOM]{electro-optic modulator}
\acrodef{fsr}[FSR]{free spectral range}
\acrodef{vco}[VCO]{voltage-controlled oscillator}
\acrodef{pbs}[PBS]{polarizing beam splitter cube}

\begin{document}

\preprint{APS/123-QED}

\title{Observation of photo-thermal feed-back in a stable dual-carrier optical spring}
%\title{Manuscript Title:\\with Forced Linebreak}% Force line breaks with \\
 \author{David Kelley}
 \email{dbkelley@syr.edu} 
\affiliation{Department of Physics, Syracuse University, \\
 Syracuse, NY, 13244 - 1130, USA }
\author{James Lough}
 \email{james.lough@aei.mpg.de}
\affiliation{Department of Physics, Syracuse University, \\
 Syracuse, NY, 13244 - 1130, USA }
\affiliation{Max-Planck-Institut f{\"u}r Gravitationsphysik (Albert-Einstein-Institut)
und Leibniz Universität,\\
Hannover, Callinstr. 38, 30167 Hannover, Germany}
 \author{Fabian Manga\~na-Sandoval}
 \email{fmaganas@syr.edu}
\affiliation{Department of Physics, Syracuse University, \\
 Syracuse, NY, 13244 - 1130, USA }
\author{Antonio Perreca}
 \email{aperreca@ligo.caltech.edu}
\affiliation{Department of Physics, Syracuse University, \\
 Syracuse, NY, 13244 - 1130, USA }
\author{Stefan W. Ballmer}
 \email{sballmer@syr.edu}
\affiliation{Department of Physics, Syracuse University, \\
 Syracuse, NY, 13244 - 1130, USA }

 %\author[1]{David Kelley}
 %\email{dbkelley@syr.edu} 
%\author[1,2]{James Lough}
 %\email{jdlough@syr.edu}
 %\author[1]{Fabian Manga\~na-Sandoval}
 %\email{fmaganas@syr.edu}
%\author[1]{Antonio Perreca}
 %\email{aperreca@syr.edu}
%\author[1]{Stefan W. Ballmer}
 %\email{sballmer@syr.edu}
%\author[1,2]{James Lough}
 %\email{jdlough@syr.edu}
%\affil[1]{%
 %Department of Physics, Syracuse University, \\
 %Syracuse, NY, 13244 - 1130, USA }
%\affil[2]{%
%Albert Einstein Institute,\\
%Hannover, Callinstr. 38, 30167 Hannover, Germany}

%\affiliation{
%\affiliation{
 %Third institution, the second for Charlie Author
%\affiliation{%


\date{\today}% It is always \today, today,

\begin{abstract}
We report on the observation of photo-thermal feed-back in a stable dual-carrier optical spring. The optical spring is realized in a 7~cm Fabry-Perot cavity comprised of a suspended 0.4~g small end mirror and a heavy input coupler, illuminated by two optical fields. The frequency, damping and stability of the optical spring resonance can be tuned by adjusting the power and detuning of the two optical fields, allowing for a precise measurement of the absorption-induced photo-thermal feed-back. The magnitude and frequency dependence of the observed photo-thermal effect are consistent with predicted corrections due to transverse thermal diffusion and coating structure. While the observed photo-thermal feed-back tends to destabilize the optical spring, we also propose a small coating modification that would change the sign of the effect, making a single-carrier stable optical spring possible.

%\begin{description}
%%\item[Usage]
%%Secondary publications and information retrieval purposes.
%42.79.Bh, 95.55.Ym, 04.80.Nn, 05.40.Ca
%%May be entered using the \verb+\pacs{#1}+ command.
%%\item[Structure]
%%You may use the \texttt{description} environment to structure your abstract;
%%use the optional argument of the \verb+\item+ command to give the category of each item. 
%\end{description}
\end{abstract}


\maketitle

\newcommand{\tcr}{\textcolor{red}}
\newcommand{\tcb}{\textcolor{blue}}
\newcommand{\tcm}{\textcolor{magenta}}
\newcommand{\tcg}{\textcolor{green}}
\newcommand{\tcp}{\textcolor{purple}}
\newcommand{\irm}{\mathrm{i}}

\newcommand{\del}[0]{{}_{{}^\triangle}\!}
\newcommand{\vk}[0]{{\bf k}}
\newcommand{\w}[0]{{\rm w}}
\newcommand{\omg}[0]{{{\Omega}}}
\newcommand{\eq}[1]{equation \ref{#1}}

%\tableofcontents

\section{Introduction}
The Advanced Laser Interferometer Gravitational-Wave Observatory (aLIGO) \cite{Harry2010}, together with its international partners Virgo \cite{2013ASPC..467..151D} and KAGRA \cite{Somiya:2011np}, aim to directly observe gravitational waves emitted by astrophysical sources such as coalescencing of black hole and neutron star binary systems. The installation 
of the Advanced LIGO detectors is completed, and commissioning towards the the first observation run is ongoing. Preliminary astrophysical data is expected in 2015. The sensitivity of those advanced gravitational-wave detectors in the observation band is limited by the quantum noise of light and the thermal noise associated with mirror coatings. A contributor to the thermal noise, expected to dominate in future cryogenic gravitational-wave detectors, is thermo-optic noise \cite{Braginsky2000303, PhysRevD.63.082003, PhysRevD.78.102003}. It is caused by dissipation through thermal diffusion.

The same physics also leads to an intensity noise coupling, known in the literature as photo-thermal effect \cite{Braginsky19991}. The low frequency behaviour of the photo-thermal effect was predicted in \cite{PhysRevD.63.082003} and experimentally measured in a Fabry-Perot cavity in by De Rosa et. al. \cite{PhysRevLett.89.237402}. The physics relevant for the the high frequency behaviour, dominated by the details of the coating, was investigated in \cite{PhysRevD.78.102003} in the context of studying thermo-optic noise. It was extended to a full model of the photo-thermal transfer function in \cite{PhysRevD.91.023010}.
Here we explore the thermo-optic effect in the context of an optical spring. The coupling acts as an additional feed-back path. The phase of the coupling becomes important and can directly affect the stability of the optical spring resonance. We can exploit this dependence for a precision measurement of the photo-thermal coupling, even if it is driven by the residual few-ppm absorption of a high-quality optic.

%photo-thermal effect in microresonators \cite{metzgercavity2004}
The desire to lower the quantum noise in the gravitational-wave observation band has driven the power circulating in the Advanced LIGO arm cavities up to about 800~kW. 
%The sensitivity of such detectors is designed to operate at the point where the quantum radiation pressure noise and the shot noise are
%at the same level in the detection band.  This sensitivity limit is commonly known as the standard quantum limit \cite{Caves80, Ni86}.
%Operating at the standard quantum limit requires high laser power, 
The high laser power, in turn, couples the angular suspension modes of the two cavity mirrors. This Sidles-Sigg instability \cite{Sidles06} creates a soft (unstable) and a hard mode, whose frequency increases with the intra-cavity power. The detector's angular control system must control the soft and damp the hard mode, and at the same time must not contaminate the observation band, starting at $10~{\rm Hz}$ in the case of Advanced LIGO. 
Future gravitational wave detectors aim to extend the observational band to even lower frequencies, further aggravating this limitation.
%Perreca et al.
We previously proposed a model \cite{Perreca14}  to overcome the angular instabilities, based on a dual-carrier optical spring scheme demonstrated by Corbitt et al., in 2007 at the LIGO laboratory \cite{Corbitt07}.
%for the length control of a cavity.
%\tcm{for a longitudinal trap of a gram-scale mirror}.
The proposed angular trap setup uses two dual-carrier beams to illuminate two suspended optical cavities which share a single end mirror. %in order to create two longitudinal trap displaced from the center of the gravity of the end mirror.
As first step towards the experimental demonstration of  the scheme we built and operated a prototype, single-cavity optical trap, capable of controlling the cavity length only \cite{LoughThesis}. The data presented in this paper was taken with this prototype.
The next version of the angular trap setup will also allow us to measure the photo-thermal effect on a folding mirror. Heinert et. al. \cite{PhysRevD.90.042001}  predicted excess thermal noise for folding mirrors due to transverse heat diffusion.
The result has not yet been experimentally confirmed, but since the same physics will also lead to an enhanced photo-thermal transfer function, 
the prediction can be verified with a photo-thermal transfer function measurement.

%A single dual beam can trap the end mirror longitudinally.  With the addition of a second dual beam, an additional degree of freedom, e.g. yaw, can be controlled. The experimental demonstration of this model is underway.
%However the proof of the angular trap with a pair of dual beams requires a longitudinal stabilization first.

%In this paper we give an experimental demonstration of the longitudinal trap using a dual-beam injected into a suspended optical cavity with a gram-scale end mirror and we set the requirements to turn off the electronic feedback in order to use the radiation pressure feedback alone. 

%As a  result we obtain the stabilization of the mirror using the radiation pressure and the electronic together. 
%In order to switch completely from the electronic feedback to the radiation feedback system, the mirror suspension should be improved by \tcm{1.314} order of magnitude in the frequency band of interest.

The paper is structured as follows: Sections \ref{sec:DCOS} and \ref{sec:PTE} will review the idea of a dual-carrier optical spring and the photo-thermal effect respectively. 
Section \ref{sec:exp} describes the experimental setup and we discuss the result in section \ref{sec:res}. Finally, section \ref{sec:SCs} suggests a coating modification to make a single-carrier optical spring feasable.
%4 reviews our experimental results.  Section 5 gives our conclusions and lays out the path for the next phase of the project, building an angular trap.}

\section{Dual-carrier optical spring}
\label{sec:DCOS}
A Fabry-Perot cavity detuned from resonance couples the intra-cavity power linearly to the mirror position. The response is delayed by the cavity storage time. The resulting optical spring constant is given by \cite{Perreca14}.
\begin{eqnarray}
\label{eq:KOS1}
K_{OS}^{\rm 1 field} & \tcb{\approx} & K_0
\frac{1}{1+\frac{\delta^2}{\gamma^2}-\frac{\Omega^2}{\gamma^2}+i2\frac{\Omega}{\gamma} } \\
K_0 &= & P_0 t_1^2 r_2^2 \frac{8k r_1r_2}{c(1-r_1r_2)^3}\frac{ \frac{\delta}{\gamma}}{(1+\frac{\delta^2}{\gamma^2})} 
\end{eqnarray}
where $P_0$ is the incident power, corrected for mode-matching losses, $k = {2\pi}/{\lambda}$ is the wave vector of the light, $t_i$ and $r_i$ are the mirror amplitude transmissivity and reflectivity for input coupler ($i=1$) and end mirror ($i=2$), and $\gamma$, $\delta$ and $\Omega$ are the cavity line, cavity detuning, and mechanical frequency. The value of $K_{OS}$ lies in either the 2nd or 4th quadrant of the complex plane, and the associated radiation pressure force creates either
a anti-restoring and damping (red detuning) or
a restoring and anti-damping force (blue detuning) \cite{Sheard04}. 

Two spatially overlapping optical fields, the carrier and sub-carrier, with opposite detuning sign and with an opportune power ratio can be used to cancel the instability \cite{Corbitt07}. The total optical spring $K_{OS}$ is the sum of the individual springs
\begin{eqnarray}
\label{eqn:KOSsum}
K_{OS}=K_{OS}^c+K_{OS}^{sc}
\end{eqnarray}
Where $K_{OS}^c$ and $K_{OS}^{sc}$ are given by equation \ref{eq:KOS1}. The dual-carrier optical spring
can be tuned to lie in the 1st quadrant for the frequency band of interest. When acting on a suspended cavity end mirror with mass $m$ and mechanical suspension spring constant $K_m$ the optical spring becomes a feed-back loop with a closed loop response function
\begin{eqnarray}
\label{eqn:TFloop}
\frac{x}{F_{ext}}=\frac{1}{-m\Omega^2+K_m+K_{OS}}
%\frac{x}{F_{ext}}=\frac{1}{-m\Omega^2+K_m+K_{OS}+K_{\rm extra}}
\end{eqnarray}
The tunability of the optical spring $K_{OS}$ in both magnitude and phase allows experimental fine-tuning of the poles of equation \ref{eqn:TFloop} to lie exactly on the real axis, resulting in an infinite Q of the optical spring (critical stablility).
Experimentally this can be done up to a maximum $Q$, above which the measured transfer function data no longer permits distinguishing between a stable and an unstable spring. The phase of the total spring constant at resonance can then be determined with a precision given by $1/Q$.
The suspension mechanical spring constant has to have a positive imaginary part, but it can be designed to be very small. Loss angles of $10^{-5}$ are easily achievable, and are further diluted by the magnitude of the ratio of $K_{OS}/K_m$. The contribution to the phase of the total spring constant from the mechanical suspension is thus expected to be negligible. The imaginary part of the optical spring $K_{OS}$ on the other hand is closely related to its real part through equations \ref{eqn:KOSsum} and \ref{eq:KOS1}, and is very accurately predicted based on the resonance frequency, carrier to sub-carrier power ratio as well as the detuning of carrier and subcarrier, i.e. only power ratios and frequencies. However, we will see below that the photo-thermal effect can affect the total transfer function. The first indication of the photo-thermal effect will be a deviation $\phi$ in phase from the expectation of equation \ref{eqn:KOSsum} around the optical spring resonance. This is easily and repeatably observable with a precision given by the inverse of the experimentally resolvable $Q$, and an accuracy determined only by frequency and power ratio measurements.
As a function of any such phase deviation $\phi$ on resonance, the optical spring closed loop response (equation \ref{eqn:TFloop}) becomes
\begin{eqnarray}
\label{eqn:TFloopPhi}
\frac{x}{F_{ext}}=\frac{1}{-m\Omega^2+(K_m+K_{OS})(1+i \phi)},
\end{eqnarray}
which on resonance is $\approx [m\Omega^2_{\rm res} i (\phi_0 + \phi)]^{-1}$. Here $\phi_0$ is the known phase of the dual optical spring from equation \ref{eqn:KOSsum} on resonance.

\section{Photo-thermal effect}
\label{sec:PTE}
\begin{figure}[ht]
\includegraphics[width=\columnwidth]{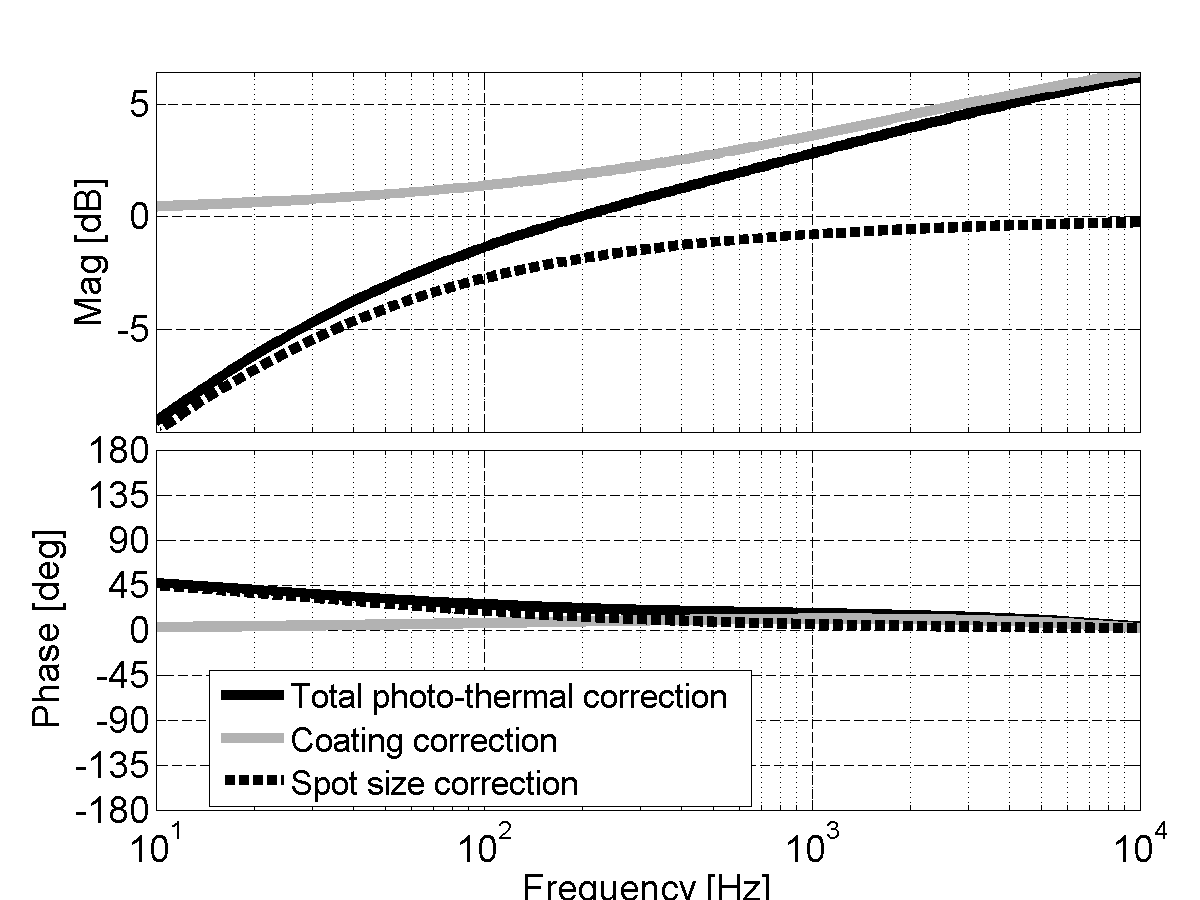}%
\caption{Correction factors for the photo-thermal transfer function of a fused silica mirror with a dielectric coating (solid black). The solid grey trace is the coating correction for a 13-doublet $\lambda/4$ $\rm Ta_2O_5\!\!:\!SiO_2$ coating. The dashed black trace shows the effect of a Gaussian beam spot with $\w=161~\mu{\rm m}$ radius. To get the full transfer function, multiply with \eq{eq:simple}, adding an overall $1/f$ shape.
The calculation is based on material parameters show in table \ref{SiO2}. 
}%
\label{fig:PTcorr}%
\end{figure}

Power absorption on the surface of an optic leads to an increase of the surface temperature. The depth of the heated layer is given by the diffusion length $d_{\rm diff}=\sqrt{\kappa/(\rho C \Omega)}$, where $\kappa$, $C$ and $\rho$ are the thermal conductivity, heat capacity and density of the material, and $\Omega$ is the observation angular frequency. In the large-spot size limit, i.e. $\w \gg d_{\rm diff}$, and neglecting coating effects,
% or $\Omega \gg \omg_c$, where $\omg_c=\kappa/(\rho C \w^2)$, 
the displacement of the surface is given by (e.g. \cite{PhysRevD.63.082003,PhysRevD.91.023010})
\begin{equation}
\label{eq:simple}
\del z = \bar{\alpha} \int_0^\infty \!\!\!\!\!\!T dz = \bar{\alpha} \frac{j}{i \omg \rho C}
\end{equation}
where $\bar{\alpha}=2(1+\sigma) \alpha$ is the effective expansion coefficient under the mechanical constraint that the heated spot is part of a much larger optic \cite{PhysRevD.78.102003,PhysRevD.70.082003}. $\alpha$ and  $\sigma$ are the regular linear expansion coefficient and Poisson ratio. $j=P/(\pi \w^2)$ is the absorbed average surface intensity of the Gaussian beam with beam radius $\w$ ($1/e^2$ intensity). This simple picture needs two important refinements. First, for frequencies  $\Omega$ around and below $\omg_c=2 \kappa/(\rho C \w^2)$ the transverse heat diffusion leads to a multiplicative correction factor to \eq{eq:simple}
derived by  Cerdonio et al. \cite{PhysRevD.63.082003}:
\begin{equation}
\label{eq:Cerdonio}
I(\omg/\omg_c) = \frac{1}{\pi} \int\limits_0^\infty du \int\limits_{-\infty}^\infty dv \frac{u^2 e^{-u^2/2}}{(u^2+v^2)\left(1+\frac{(u^2+v^2)}{i \omg/\omg_c} \right) }
\end{equation}
As expected, for $\omg \gg \omg_c$, the correction factor approches 1. 
%which takes a simple form in the transverse Fourier space.
%If this displacement is read out by the same beam profile as the heating intensity (normalized intensity $\hat{I}$), the effective displacement becomes
%\begin{equation}
%\del z_{\rm eff} =  \alpha (1+\sigma) \int d^2 \vk \frac{I_\vk \hat{I}^*_\vk }{i \Omega \rho C+ \kappa \vk^2} =\frac{\alpha (1+\sigma)}{i \Omega \rho C}  \int d^2 \vk W(\vk) \frac{i \frac{\Omega}{\Omega_\vk}  }{1 + i \frac{\Omega}{\Omega_\vk}}
%\label{eq:zeff}
%\end{equation}
%where $\Omega_\vk=\frac{\kappa \vk^2}{\rho C}$, and a weighting function $W(\vk)$ obeying $\int d^2 \vk W(\vk) = I$, the mean intensity, such that we recover equation \ref{eq:simple} in the high frequency limit. Equation \ref{eq:zeff} implies that in general the transfer function from absorbed beam intensity to displacement is given by a weighted sum of poles, where the weighting function is given by the spatial beam profile. For simplicity the derivation given here neglects corrections due to the underlying elasticity problem. This is done in \cite{PhysRevD.63.082003} and results in a slight change in the weighting function $W(\vk)$. 
For a fused Silica substrate, $\rm SiO_2$,  and a Gaussian beam spot radius of $\w=161~\mu{\rm m}$ this correction becomes large below $\omg_c/(2 \pi) = 10~{\rm Hz}$, but is measurably different from unity even at $1~{\rm kHz}$. (See fig \ref{fig:PTcorr})

Second, for high frequencies, the diffusion length becomes comparable to the coating thickness. Since the optical field is reflected by a dielectric stack, the effective mirror displacement is given by \cite{PhysRevD.78.102003,PhysRevD.91.023010}
\begin{equation}
\label{eq:dphic2}
\del z =  \sum_{i}   \left[ \frac{\partial \phi_{\rm c}}{\partial \phi_i} (\beta_i \!+\! \bar{\alpha}_i n_i) 
\!+\!  \bar{\alpha}_i  \right]  \bar{T}_i d_i
\end{equation}
where  $\bar{\alpha}_i$, $\beta_i=dn/dT$ and $n_i$ are the constrained effective expansion coefficient, the temperature dependence of the index of refraction, and the index of refraction itself for layer $i$. $\frac{\partial \phi_{\rm c}}{\partial \phi_i}$, the dependence of the coating reflected phase on the round trip optical phase in layer $i$, is always negative, resulting in a sign change and enhancement of the bracket in \eq{eq:dphic2} for the first few layers. $\bar{T}_i d_i$ is the temperature profile driven by the absorped intensity $j$, integrated across layer $i$. For a $\rm Ta_2O_5\!\!:\!SiO_2$ coating used in gravitational wave detectors we find a measureable enhancement of the photo-thermal transfer function around $1~{\rm kHz}$ \cite{PhysRevD.91.023010}. Additionally, depending on the detailed absorption profile, a sign change can occur above about $100~{\rm kHz}$.

For the experiment parameters discussed in this paper, i.e. a  Gaussian beam spot radius of $\w=161~\mu{\rm m}$ and a mirror coating with about 13 doublet layers both effects are relevant in the $100~{\rm Hz}$ to $1~{\rm kHz}$ band. Their contributions are plotted in figure \ref{fig:PTcorr}.

\section{Experimental setup}
\label{sec:exp}

\subsection{Cavity}

%\begin{table}[h]
%\scriptsize
%\begin{tabular}{|l|l|l|l|l|l|l|}
%\hline
%$\lambda_0$ & Mirror RoC & $L_0$    & Spot size  & FSR      & Finesse & Cavity Pole \\ \hline
%1064 nm     & 5.0 cm     & 7.0 cm & 161 $\mu$m & 2.14 GHz & 7500    & 143 KHz     \\ \hline
%\end{tabular}
%%\end{table}
%%
%%\begin{table}[h]
%\begin{tabular}{|l|l|l|l|}
%\hline
%$\delta f_{C}$ & $\delta f_{SC}$ & $P_{C}$    & $P_{SC}$ \\ \hline
%213-290 KHz    & 27-36 KHz     & 225-239 mW & 65-78 mW \\ \hline
%\end{tabular}
%\end{table}

\begin{table}[h]
\begin{tabular}{|l|l|}
\hline
$\lambda_0$ & 1064 nm \\ \hline
Mirror RoC & 5.0 cm \\ \hline
$L_0$ & 7.0 cm \\ \hline
Spot size  & 161 $\mu$m\\ \hline
FSR      & 2.14 GHz \\ \hline
Finesse & 7500 \\ \hline
Cavity Pole & 143 KHz\\ \hline
\end{tabular}
%\end{table}
%
%\begin{table}[h]
\begin{tabular}{|l|l|}
\hline
$\delta f_{C}$ & 213-290 KHz \\ \hline
$\delta f_{SC}$ & 27-36 KHz \\ \hline
$P_{C}$ input& 225-239 mW \\ \hline
$P_{SC}$ input & 65-78 mW \\ \hline
\end{tabular}
\caption{Parameters of the optical spring cavity. The range of values for the carrier and sub-carrier detuning frequency ($\delta f_{C}$, $\delta f_{SC}$) and input power ($P_{C}$, $P_{SC}$) indicate the variation between individual measurements.}
\label{tab:params}
\end{table}

\begin{figure}[ht]
\includegraphics[width=.8\columnwidth]{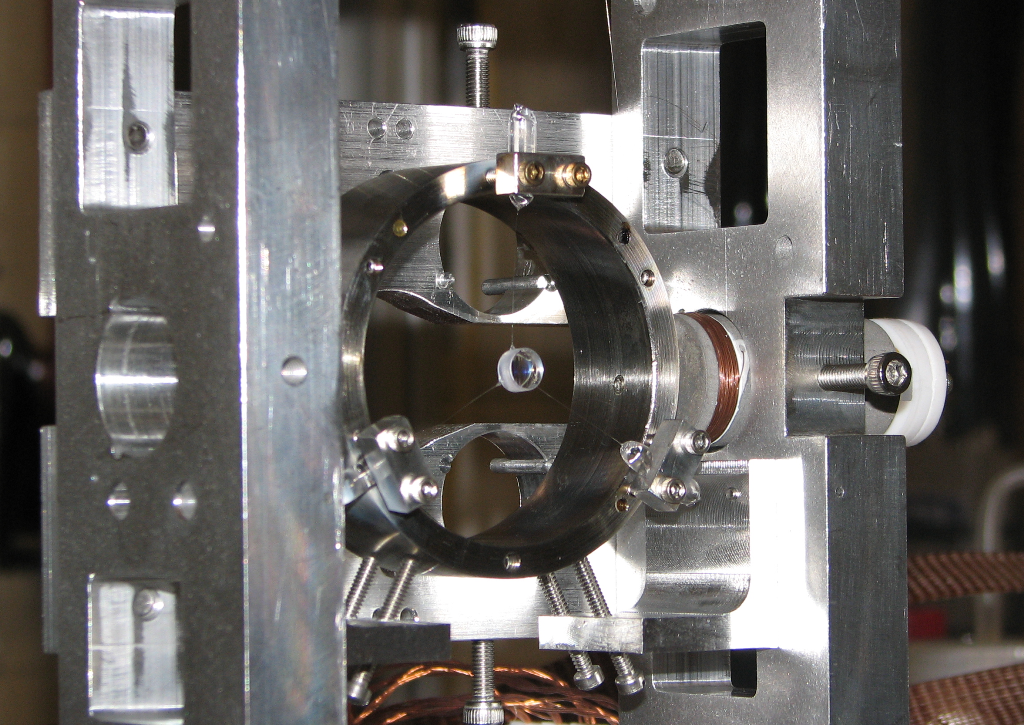}%
\caption{A picture of the small end mirror suspended from a steel ring by glass fibers. The ring is suspended from a small optics suspension (SOS) with tungsten wire.
The SOS provides DC alignment control while allowing the mirror to move freely above the 18Hz resonance of the fiber suspension. The end of the fiber is a small glass nub attached to the mirror with epoxy. This produces a fairly high suspension Q of about $5 \cdot 10^5$. The resulting contribution of damping in the opto-mechanical spring is insignificant compared to the damping from the optical field.}%
\label{fig:smallmirrorpic}%
\end{figure}

The optical spring cavity is composed of two suspended mirrors in a vacuum chamber, each with radius of curvature RoC = 5$\,$cm and power transmissivity $T = 4.18\times10^{-4}$.
The measured finesse is $7500\pm 250$ and the cavity length is $L_0 = 7.0\pm0.2\,$cm. We chose a short cavity to minimize frequency noise coupling. The cavity has a free spectral range (FSR) of about 2.14$\,$GHz and cavity pole $f_{pole} = \gamma/(2 \pi) = 143~{\rm kHz}$. 
The input mirror mass is $300\,$g, designed to be heavy to make it insensitive to radiation pressure; it is suspended as a single stage pendulum with mechanical resonances, i.e. position, pitch and yaw, close to $1\,$Hz. The end mirror has a mass of $0.41\pm 0.01~{\rm g}$ and is $7.75~{\rm mm}$ in diameter. It is suspended with three glass fibers from a $300~{\rm g}$ steel ring, shown in figure \ref{fig:smallmirrorpic}. The steel ring has diameter of $7.6~{\rm cm}$ and is itself suspended.
% and damped by local active feedback.
%described in Sec.$\,$\ref{sec:glasssus}. 
%The glass fiber resonances are about 18 Hz. 
The input mirror is actively controlled by an electronic feedback system, while the end mirror is 
free to move in the glass suspension abobe its resonance frequency of $18~{\rm Hz}$, and is only subject to the optical spring radiation pressure. 
%The end mirror motion is locked to the input mirror using optical springs and laser feedback.

\subsection{Input field preparation}
\label{sec:layout}

\begin{figure}[htb]%
\centering
\includegraphics[width=\columnwidth]{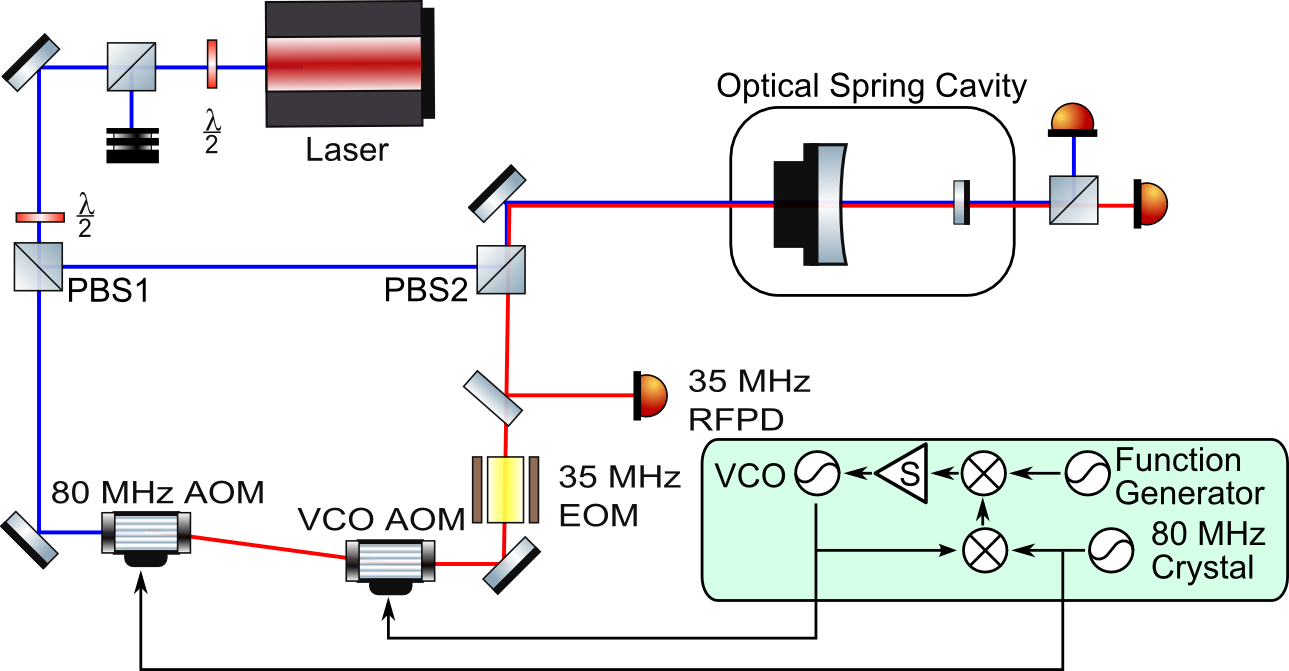}%
\caption{A schematic layout of the optical trap experiment. The light from the laser is split into the carrier and subcarrier paths at PBS1, with a ratio determined by the $\lambda/2$ plate. The subcarrier path is frequency shifted by two AOMs under the control of the subcarrier servo (described in detail in Section \ref{sec:subservo}), then recombined with the carrier at PBS2. The co-aligned mode-matched beams enter the cavity, then are individually monitored at the output. We can use the 35 MHz EOM and RFPD in a PDH scheme to read out the cavity length or lock the cavity.}
\label{fig:layout}
\end{figure}

%The experimental setup is shown in figure \ref{fig:layout}. 
The optical field incident on the optical spring cavity consists of two beams, a carrier and a subcarrier, as described in Section \ref{sec:DCOS}.
As shown in figure \ref{fig:layout}, a 1064 nm laser is split into a carrier and a subcarrier beam at the polarizing beam splitter PBS1. 
In the subcarrier path two acoustic optic modulators (AOMs) are used to impose a relative frequency shift $\Delta$, 
on the subcarrier beam, leaving it at a set detuning from the carrier beam.  
$\Delta$ is set using an external signal generator (see Sec.\,\ref{sec:subservo}).
The two beams recombine at PBS2 and proceed towards the Fabry-Perot cavity with opposite polarization.
The total power and the power ratio between the carrier and subcarrier beams are set by two half wave-plates $\lambda /2$. 

The subcarrier beam is modulated by a 35 MHz electro-optic modulator (EOM). We measure the modulated light reflected by the cavity with a resonant radio-frequency photodiode (RFPD) and then demodulate to read out the cavity length with the Pound-Drever-Hall technique (PDH) \cite{Black01}.  We use the subcarrier to derive a PDH singal because the subcarrier requires less detuning than the carrier. We can use the PDH signal to actuate on the laser and the suspensions to lock the cavity, then turn down the gain and use the PDH signal for readout. 

A small offset added to the PDH error signal shifts the locking point of the cavity to the side of the resonance, setting the subcarrier detuning $\delta_{sc}$. 
We choose to introduce an offset that corresponds to a negative frequency (``red'') detuning. Consequently the carrier is positively (``blue'') detuned at  
$\delta_c = \Delta + \delta_{sc}$. An electronic locking servo can be used to process the error signal and feed back to coils, actuating on
magnets mounted on the large cavity mirror, and to the laser frequency.
%the control signal at the laser in order to have a stable controlled system.

\subsection{Subcarrier Servo}
\label{sec:subservo}

The high FSR of our cavity (2.14 GHz) meant that available AOMs, with much lower operating frequency ranges (65 to 95$\,$MHz), were not suitable to lock the carrier and subcarrier on adjacent resonances.  
However, this same operating range prevents a single AOM from locking the two beams on the same resonance, due to the small cavity linewidth.
%
%we must operate with the two beams (carrier and subcarrier) detuned around the same resonance
%as the minimum drive frequency is higher than the linewidth, while 
%the maximum is less than the \ac{fsr}.
%
Thus, we set the subcarrier on the same resonance fringe as the carrier using two AOMs, each one shifting the laser frequency by about 80MHz in opposite directions.
One is driven by an $80~{\rm MHz}$ crystal oscillator, while the other is driven by a servo-locked Voltage controlled oscillator (VCO) running slightly offset from $80~{\rm MHz}$ (see figure \ref{fig:layout}).
%, which gives us the
%knob to detune the subcarrier relative to the carrier.
%{Figure \ref{fig:layout} shows the basic layout of the subcarrier servo loop.
To control the offset frequency the $80~{\rm MHz}$ signal from the crystal oscillator is mixed with the VCO output, producing a signal at the frequency difference. This difference signal is then mixed with the drive from a function generator, creating the error signal for the servo.  The servo drives the frequency modulation input of the VCO, closing the loop and locking the subcarrier beam to a fixed frequency offset from the carrier beam.

%\tcg{We produce a beat signal between the two oscillators and we lock the beat signal to a function generator which is set at our desired frequency offset. 
%Thus we can set the carrier-to-subcarrier offset frequency directly using the knob on the function generator.} \ \tcm{Not clear to me.} \tcm{We may need a pic of the electronic setup}
This setup significantly suppresses the frequency noise from the VCO. The remaining subcarrier frequency noise (relative to the carrier) is dominated by fluctuations in the path length difference between carrier and sub-carrier, see figure \ref{fig:layout}.

\input{OT_longit_V2_part2}

%\section{References}
\bibliographystyle{iopart-num}
%\bibliography{test2.bib}
\bibliography{OT_paper_long}

\end{document}

%% file: OT_longit_V2_part2.tex
\section{Results}
\label{sec:res}

\begin{figure*}[thb]%
\centering
\includegraphics[width=.7\paperwidth]{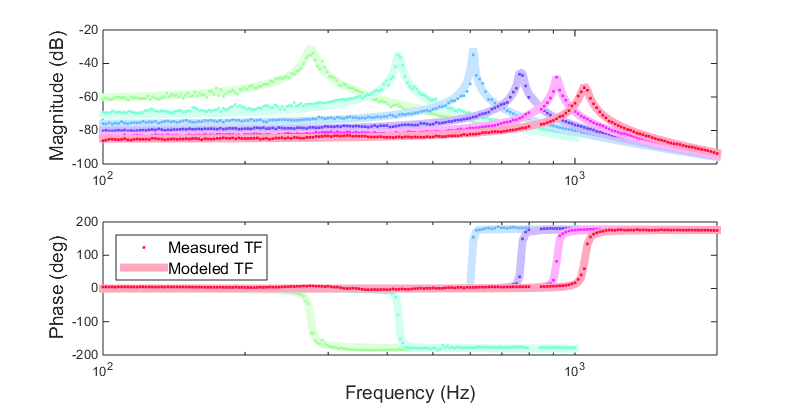}%
\caption{Data and modeled transfer function for a series of stable and unstable springs. The modeled transfer functions include the full coating and spot size correction, computed with the measured average absorption. Stable springs show a phase drop of 180 degrees at resonance, while unstable springs show a rise of 180 degrees. The magnitude is given in dB meter/Newton.}%
\label{fig:springs}%
\end{figure*}
%\tcm{MUST BE REWRITTEN IN A POSITIVE WAY. WE have nice results to show not only problems. Because the previous version was literally depressing
%I remove it. Jim this is a section that you should have in your thesis, so you can plug it in and we will see how it flows.}
%\tcm{We also need to to say what function we are plotting...equation 1? 2? }

%At the beginning of the experiment, our goals were to measure and then accurately model a series of stable and unstable optical springs. During the course of the measurements, we saw a consistent trend towards instability; each measurement we took was more unstable than our model predicted it should be. We realized that this behavior was exactly what we would see from the photo-thermal effect. We made a new goal of measuring the effects of the photo-thermal absorption.

Using the setup described in the previous section, we locked the cavity using a PDH error signal from the sub-carrier, feeding back to the laser frequency actuator and, at low frequencies, the heavy input coupler position. The unity gain frequency was $20~{\rm kHz}$, while the cross-over frequency between laser frequency and  input coupler position actuation was $250~{\rm Hz}$. In this configuration we fine-tuned the optical spring parameters (carrier and sub-carrier offset and power) and 
measured the PDH control loop open loop transfer function. Dividing out the known PDH loop sensing and actuation function gives us the closed loop transfer functions of the optical springs (figure \ref{fig:springs}). While we demonstrated stable and unstable dual-carrier optical springs, these measurements revealed a significantly smaller phase margin of the optical spring than expected based on equation \ref{eqn:TFloop}, suggesting the presence of a non-radiation-pressure feed-back path.

At a few ppm, the absorption $A$ of the mirrors has a very small effect on the cavity finesse and no significant impact on the total transmitted power. However, this small amount of absorption still causes local heating of the optic, driving fluctuations in the surface position of the optic, and thus the cavity length, via the photo-thermal effect. If this is the dominant effect, we should be able to include the photo-thermal effect in our model and fit the model to the data, using the absorption as the free parameter. Given a set of optical spring measurements done under similar conditions, we would then expect to find a consistent absorption coefficient across measurements.

\subsection{Analysis}

For each measured optical spring transfer function we record the carrier and subcarrier transmitted powers, $P_{tc}$ and $P_{ts}$, the optical spring resonance frequency $f_{res}$, and the difference between the carrier and subcarrier detunings $df_c-df_s$, which is set by the function generator frequency.   

%Using these values we can accurately determine the parameters [$P_{is}$,$P_{ic}$, $df_c$,$df_s$], namely
%the power coupling into the cavity and the detuning for both the carrier and the subcarrier at the time of measurement. %\tcr{This is useful because we had a little bit of alignment drift during the course of the experiment.}
%Equation \ref{eqn:TFloop} allows us to convert these four parameters into an optical spring transfer function.

We can then fit the data $d$ using a model $m$, which includes the photo-thermal effect. In particular we fit the ratio $d/m$ using a least-squares fit to minimize $E$, the error.
\begin{equation}
E=\Sigma \left|\frac{d}{m}-1\right|^2 
\end{equation}

We fit for a small magnitude offset, the subcarrier detuning $df_s$, and the absorption $A$. We assess the fitting errors by modeling the noise in each frequency bin of the transfer function measurement, and propagating this noise through the fit. Four of the optical spring transfer functions  had a measurement noise of a little less than $1~{\rm dB}$, while the optical springs at $276~{\rm Hz}$ and $422~{\rm Hz}$ had a significantly higher noise of about $3~{\rm dB}$. We think this noise is dominated by intra-cavity power fluctuations, most likely due to angular fluctuations.

The remaining parameters (cavity transmitted powers and carrier-sub-carrier frequency spacing) we treat as systematic errors. We propagated their measurment errors through the fit. We used a $2\%$ measurement error for the power measurements and  a $1~{\rm kHz}$ error for the frequency separation.

\begin{figure}[htb]%
\includegraphics[width=\columnwidth]{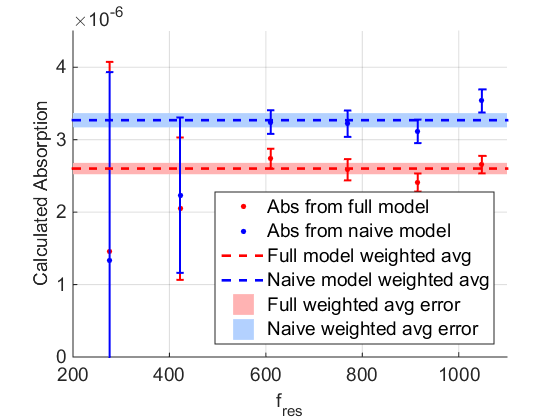}%
\caption{Absorption fit for naive and full models. The full model absorption is consistent with a constant absorption of  $2.60\pm0.08$ ppm. The naive $1/f$ model predicts $3.27\pm0.10$ ppm. The transfer function data for the lowest two resonant frequencies was significantly noisier. Also, at lower frequencies the photo-thermal  effect has a smaller effect on the total optical spring. Both effects result in the larger error bars at low frequencies.}
\label{fig:abs}%
\end{figure}

After determining the absorption $A$ for each optical spring transfer function measurement, we can take a statistical-error-weighted average to arrive at the most probable absorption coefficient for the mirror.  For the full photo-thermal model we measure a consistent absorption of  $2.60\pm0.08$ ppm ($\pm 0.06$ ppm statistical,  $\pm 0.05$ ppm systematic) (see figure \ref{fig:abs}). The naive $1/f$ model yields an absorption of $3.27\pm0.10$ ppm ($\pm 0.08$ ppm statistical,  $\pm 0.06$ ppm systematic).  The detailed model with coating and spot size corrections is slightly preferred by the data over the naive $1/f$ model, i.e. the result is more consistent with the same absorption at all frequencies. However the errors in our mesurement are too large to make this statement with any sigificant certainty.

%\begin{eqnarray}
%A_ = \frac{\Sigma\left(A\epsilon_{A}^{-2}\right)}{\Sigma\left(\epsilon_{A}^{-2}\right)}
%\hspace{20pt}
%\sigma_{ABS} = \sqrt{\frac{1}{\Sigma\left(\epsilon_{Ai}^{-2}\right)}};
%\label{eq:weighting}
%\end{eqnarray}

\begin{figure}[htb]%
\includegraphics[width=\columnwidth]{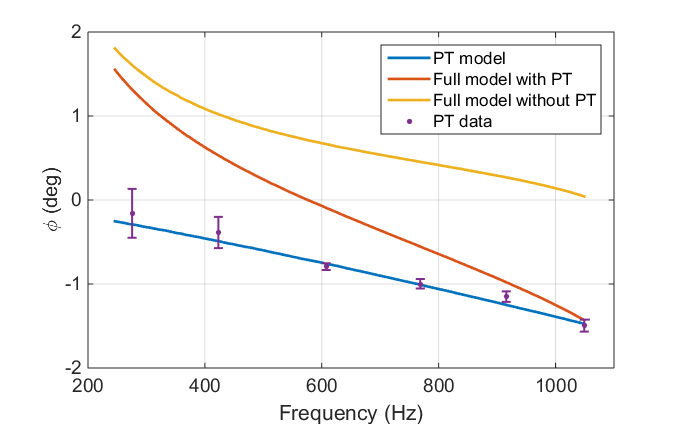}%
\caption{Feedback phase at the optical spring resonance frequency due to the optical spring and photo-thermal (PT) effect. The measured extra phase is consistent with $2.60$ ppm of absorption. The error bars are as small as $\pm 0.04^{\circ}$, a remarkable precision for an open loop transfer function phase measurement.
}
\label{fig:phi}%
\end{figure}

Since this measurement is based on the missing optical spring phase on resonance (see equation \ref{eqn:TFloopPhi}), we can also express the result as extra phase. Near the resonance the optical spring constant is close to real, while the photo-thermal effect is almost purely imaginary. Thus we approximately find for the extra phase $\phi$
\begin{equation}
\phi =  2 m \Omega^2 \frac{c}{2} \frac{\bar{\alpha}}{\Omega \rho C \w^2 \pi} A I_{\rm corr}
\approx 0.4^{\circ} \frac{A I_{\rm corr}}{1~{\rm ppm}} \frac{f}{1~{\rm kHz}} 
\label{eq:phase}
\end{equation}
Here the leading factor of two accounts for the two mirrors, $I_{\rm corr}$ is the real part of the total correction factor plotted in figure \ref{fig:PTcorr}, and we used the material parameters for fused silica (see table \ref{SiO2}).  
Figure \ref{fig:phi} shows the measured extra phase at the resonance frequency of the optical spring, together with the prediction from the photo-thermal feed-back with the best-fit absorption. The figure also shows the expected phase due to the dual-carrier optical spring, as well as the total phase of the complete model. Finally it is worth mentioning that this is a remarkably precise way to measure the phase of the open loop transfer function - the  error bars in figure \ref{fig:phi} are as small as $0.04^{\circ}$. While it is possible to measure the frequency-dependent photo-thermal  phase loss directly in a cavity held on resonance, it would be challenging to achieve the same precision. The magnitude of the effect could be increased using e.g. an external C02 laser 
to heat the surface instead of relying on residual absorption, but this would introduce subtle differences due to the different heat deposition depth and uncertainties in the beam overlap between heating and readout beam.

\section{Stable single-carrier optical spring}
\label{sec:SCs}
\begin{figure*}[thb]
\centering
\includegraphics[width=.7\paperwidth]{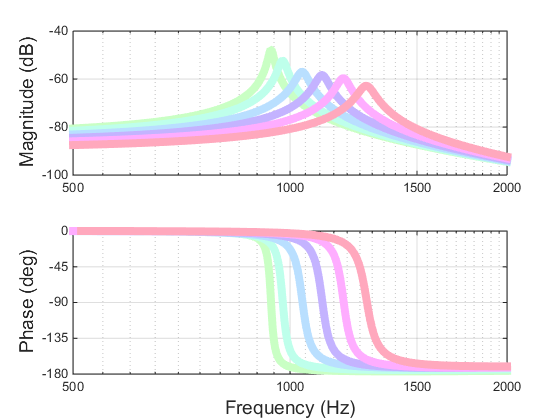}
\caption{Stable single-carrier optical springs (no sub-carrier) with modified coating - the first coating layer is $20.5$ wavelength thick. See text for details. The six traces otherwise have the same parameters as the best-fit optical springs in figure \ref{fig:springs}. The magnitude is given in dB meter/Newton.}
\label{fig:SCsprings}
\end{figure*}

In the experiment at hand the photo-thermal feed-back always pushed the optical spring resonance closer to instability.
Perhaps the most interesting question is whether we can change the sign of this feed-back path and exploit it to stabilize an otherwise unstable optical spring. It was pointed out in \cite{PhysRevD.91.023010} that this naturally occurs above about $100~{\rm kHz}$ for a regular dielectric coating. 
At those frequencies the thermal diffusion length only affects the first few layers of the coating, which affect the overall coating reflected phase differently than the rest of the coating.
However it is actually quite simple to get this sign inversion to occur at a much lower frequency. Increasing the thickness of the inital half-wavelength $SiO_2$ layer - but keeping it an odd multiple of half the wavelength - will boost the effect from the first layer, thus lowering the frequency at which this sign inversion occurs. Indeed this effect can be strong enough that the damping effect from the sub-carrier is not needed to generate a stable optical spring. To illustrate this, figure \ref{fig:SCsprings} shows a set of six optical springs with parameters identical to the ones shown in figure \ref{fig:springs}, except that we set the sub-carrier power to zero (i.e. they are single-carrier optical springs), and we increased the first $SiO_2$ coating layer from $0.5$ wavelength to $20.5$ wavelength.

Such a modified coating would thus allow detuned self-locking  of an optical cavity, using just one laser frequency. It does rely on a small amount (order 1 ppm) of optical absorption in the coating, but this level of absorption is often unavoidable anyway, and does not prevent high-finesse cavities. 

Due to their intrinsic simplicity stable single-carrier optical springs might open up a number of new applications reaching beyond their use in gravitational wave detectors. In particular the prospect of tuning feed-back in opto-mechanical applications by designing an appropriate coating is promising and might be useful in a variety of sensor applications.
Because stable single-carrier optical springs rely on optical absorption, they will allow vacuum fluctuations to enter the system, particularly at the optical spring resonance frequency. This could in principle constrain their use in quantum-limited systems. However, as in the case of our experiments, in practice the required absorption can be so small that the optical properties of the system are not limited by them.

\section{Conclusions}
We observed photo-thermal feedback in an experimental optical spring setup for a 0.4 gram mirror. We made measurements for a range of optical spring resonant frequencies, and used a least squares fit to calculate the absorption. The data is consistent with the predictions of the complete model presented in Section \ref{sec:PTE}, but only sligthly prefers it over a simple model that ignores any heat diffusion in the coating and transverse to the optical axis. We also show that a small modification of the first layer of the high-reflectivity coating would be enough to reverse the sign of the photo-thermal feed-back, to the extent that a single-carrier, dynamically and statically stable optical spring becomes feasable.

%We presented a model for a complete photo-thermal effect in a cavity optical springs. This model included radial diffusion behavior, coating expansion, and bulk expansion. We described the optical trap experiment at Syracuse University as it relates to the creation of optical springs influenced by the photo-thermal effect. We measured stable and unstable optical springs, which showed more instability than the basic optical spring model predicted. We used a naive model and a complete model of the photo-thermal effect to fit the data. We found excellent agreement between the measured data and the complete photo-thermal effect model.

Repeating the presented measurement with a folding mirror in a cavity should also allow us to confirm the predicted enhancement of thermal noise for folding mirrors \cite{PhysRevD.90.042001} . This noise will affect any gravitational-wave interferometer design making use of folding mirrors in the arm cavities \cite{Ballmer13}.

\input{tableSiO2Ta2O5}

\section{Acknowledgments}

This work was supported by the National Science Foundation grants PHY-1068809 and PHY-1352511. The authors used computer resources supported by NSF grants PHY-1040231, PHY-1104371, and PHY-0600953.
This document has been assigned the LIGO Laboratory document number LIGO-P1500003.

%% file: tableSiO2Ta2O5.tex
\begin{table}[h]
\begin{tabular}{llrrl}
Parameters ${\rm Ta_2O_5\!:\!SiO_2}$  & Symbol                   & ${\rm SiO_2}$    & ${\rm Ta_2O_5}$     & Unit                   \\
\hline
Refractive Index (@1064 nm) & $n$                         & 1.45                      & 2.06 & -                      \\
Specific Heat                               & $C$                        & 746                       & 306 & J/kg/K                 \\
Density                                         & $\rho$                    & 2200                      & 6850 & kg/m${}^3$ \\
Thermal Conductivity                 & $\kappa$               & 1.38                      &  33  & W/m/K                  \\
Thermal expansion coef.           & $\alpha$                & 0.51                      & 3.6 & ppm/K                  \\
Thermo-Optic coef.  ($\rm 1{\mu}m$)& $\beta = \frac{dn}{dT}$ & 8             & 14 & ppm/K                  \\
Poisson ratio                                & $\sigma$               & 0.17                       &  0.23 & -                 \\
Young’s Modulus                        & $E$                        & 72.80                     & 140 & GPa
\end{tabular}
\caption{Parameters for fused silica (${\rm SiO_2}$) and tantulum-pentoxide (${\rm Ta_2O_5}$). The values are taken from \cite{PhysRevD.78.102003} and \cite{PhysRevD.70.082003}.  }
\label{SiO2}
\end{table}